\documentclass[preprint,aps,pra,showpacs,floatfix]{revtex4}
\usepackage{graphicx}
\usepackage{times}
\usepackage{nicefrac}
\usepackage{amsmath}
\usepackage{amsfonts}
\usepackage{amssymb}
\usepackage{amsthm}
\usepackage{epsf}
\usepackage{bm}
\usepackage{bbm}
\usepackage{times}

\usepackage{dcolumn}
\newcolumntype{.}{D{x}{}{-1}}
\newcommand{\balpha}{{\mbox{\boldmath$\alpha$}}}
\newcommand{\bnabla}{{\mbox{\boldmath$\nabla$}}}

\newcommand{\be}{\begin{eqnarray}}
\newcommand{\ee}{\end{eqnarray}}
\newcommand{\la}{\langle}
\newcommand{\ra}{\rangle}

\newcommand{\rmd}{{\rm d}}

\newcommand{\bfp}{{\bf p}}
\newcommand{\bfr}{{\bf r}}
\newcommand{\bfx}{{\bf x}}

\newcommand{\veps}{\varepsilon}

\newcommand{\calH}{{\cal H}}

\newcommand{\bcalH}{{\mbox{\boldmath$\cal H$}}}

\newcommand{\cpsi}{\psi^{\dagger}}
\newcommand{\Epr}{E^{\prime}}
\newcommand{\tpr}{t^{\prime}}
\newcommand{\bfxpr}{\bfx^{\prime}}
\newcommand{\bra}[1]{\langle#1|}
\newcommand{\ket}[1]{|#1\rangle}

\newcommand{\matrixel}[3]{\langle #1 | #2 | #3 \rangle}

\newcommand{\ddo}{\frac{\rmd}{\rmd \omega}}
\newcommand{\dddo}{\frac{\rmd^2}{\rmd^2 \omega}}

\newcommand{\aZ}{\alpha Z}
\newcommand{\intinff}{\int_{-\infty}^{+\infty}}
\newcommand{\intinf}{\int}
\newcommand{\intG}{\oint_{\Gamma}}
\newcommand{\tomega}{\tilde\omega}
\newcommand{\gdirac}{g_{\rm D}}
\newcommand{\dgint}{\Delta g_{\rm int}}
\newcommand{\dgqed}{\Delta g_{\rm QED}}
\newcommand{\dgsqed}{\Delta g_{\rm SQED}}
\newcommand{\dgnuc}{\Delta g_{\rm nuc}}

\newcommand{\muB}{\mu_0}
\newcommand{\muN}{\mu_N}
\newcommand{\dEScrSE}{\Delta E_{\rm SQED}^{\rm SE}}
\newcommand{\dEScrSEX}[1]{\Delta E_{\rm SQED}^{{\rm SE}(#1)}}

\newcommand{\dEA}{\dEScrSEX{A}}
\newcommand{\dEB}{\dEScrSEX{B}}
\newcommand{\dEC}{\dEScrSEX{C}}
\newcommand{\dED}{\dEScrSEX{D}}
\newcommand{\dEE}{\dEScrSEX{E}}
\newcommand{\dEF}{\dEScrSEX{F}}
\newcommand{\dEG}{\dEScrSEX{G}}
\newcommand{\dEH}{\dEScrSEX{H}}
\newcommand{\dEI}{\dEScrSEX{I}}
\newcommand{\dEAi}[1]{\dEScrSEX{A#1}}
\newcommand{\dECi}[1]{\dEScrSEX{C#1}}
\newcommand{\dEEi}[1]{\dEScrSEX{E#1}}
\newcommand{\dEGi}[1]{\dEScrSEX{G#1}}
\newcommand{\dEHi}[1]{\dEScrSEX{H#1}}
\newcommand{\dEIi}[1]{\dEScrSEX{I#1}}
\newcommand{\dEScrVP}{\Delta E_{\rm SQED}^{\rm VP}}
\newcommand{\dEScrVPX}[1]{\Delta E_{\rm SQED}^{{\rm VP}(#1)}}

\newcommand{\UVP}{U_{\rm VP}^{\rm el}}
\newcommand{\UVPml}{U_{\rm VP}^{\rm ml}}
\newcommand{\IVP}{I_{\rm VP}}
\newcommand{\IVPml}{I_{\rm VP}^{{\rm ml}}}

\newcommand{\dEScrSEVP}{\Delta E_{\rm SQED}^{\rm SE/VP}}
\newcommand{\dgScrSE}{\Delta g_{\rm SQED}^{\rm SE}}
\newcommand{\dgScrVP}{\Delta g_{\rm SQED}^{\rm VP}}
\newcommand{\dgScrSEVP}{\Delta g_{\rm SQED}^{\rm SE/VP}}
\newcommand{\xScrSE}{x_{\rm SQED}^{\rm SE}}
\newcommand{\xScrVP}{x_{\rm SQED}^{\rm VP}}
\newcommand{\xScrQED}{x_{\rm SQED}}
\newcommand{\xScrSEVP}{x_{\rm SQED}^{\rm SE/VP}}
\newcommand{\xQED}{x_{\rm QED}}
\newcommand{\xSQED}{x_{\rm SQED}}
\newcommand{\vepsn}[1]{\veps_{n_{#1}}}
\newcommand{\DeltaQbPb}{\Delta}
\begin{document}

\title{
Evaluation of the
screened QED corrections to the $\bm{g}$ factor
and the hyperfine splitting of lithiumlike ions}
\author{
D.~A.~Glazov,$^{1,2}$ A.~V.~Volotka,$^{1,2}$
V.~M.~Shabaev,$^{1}$ I.~I.~Tupitsyn,$^{1}$ and G.~Plunien$^{2}$ }
\affiliation{
$^1$
Department of Physics, St. Petersburg State University,
Oulianovskaya 1, Petrodvorets, St. Petersburg 198504, Russia \\
$^2$ Institut f\"ur Theoretische Physik, Technische Universit\"at Dresden,
Mommsenstra{\ss}e 13, D-01062 Dresden, Germany \\
}
\begin{abstract}
The screened QED corrections of the first orders in $\alpha$ and $1/Z$
to the $g$ factor and the hyperfine splitting of lithiumlike ions
are evaluated within {\it ab initio} quantum electrodynamical approach.
The complete gauge-invariant set of the two-electron self-energy diagrams
in the presence of the magnetic field and a dominant part of the two-electron
vacuum-polarization diagrams are calculated.
The most accurate values of the $g$ factor of Li-like lead and uranium
are presented. The theoretical prediction for the specific difference
of the hyperfine splittings of H- and Li-like bismuth is improved.
\end{abstract}

\pacs{12.20.Ds, 31.30.Jv, 31.30.Gs}

\maketitle

%
\section{Introduction}
\label{sec:intro}
%

The bound-electron $g$ factor has been the subject of intense
experimental and theoretical investigations over the past decade.
Recent measurements for low-$Z$ hydrogenlike ions with a spinless nucleus
have reached the precision of $10^{-9}$
\cite{hermanspahn:00:prl,haeffner:00:prl,verdu:04:prl}.
Together with the corresponding theoretical studies
these experiments have lead to the new value of the electron mass,
four times more accurate than the previously accepted value
(see Ref. \cite{CODATA06} and references therein).
Experimental investigations of ions with more than one electron are anticipated
in the nearest future. In particular, measurements of the $g$ factor
of H-like and Li-like calcium and silicon are currently in progress
by the Mainz-GSI collaboration \cite{vogel:08:epj}.
An extension of these studies to high-$Z$ H-like, Li-like, and B-like systems
planned in the framework of the HITRAP project \cite{kluge:08:aqc,vogel:09:jpb}
will provide a stringent test of the bound-state QED in the strong electric field
of the nucleus.
Moreover, investigations of the $g$ factor of heavy B-like ions can lead
to an independent determination of the fine structure constant \cite{shabaev:06:prl}.
The motivation for studying the $g$ factor of Li-like and B-like ions
follows from the higher theoretical accuracy that can be reached for a specific
difference of the $g$ factor values of H-like and Li-like ions
(or H-like and B-like ions) of the same isotope.
Various effects on the $g$ factor of H-like ions were investigated
during the last two decades: one-loop \cite{blundell:97:pra,persson:97:pra,beier:00:pra,
karshenboim:01:cjp,karshenboim:02:plb,yerokhin:04:pra,lee:05:pra,yerokhin:10:pra}
and two-loop \cite{pachucki:05:pra,jentschura:09:pra} QED corrections,
recoil corrections \cite{shabaev:01:pra,shabaev:02:prl,pachucki:08:pra},
nuclear polarization effect \cite{nefiodov:02:prl}, and nuclear size effect
\cite{glazov:02:pla}.
The theoretical investigations of the $g$ factor of Li-like ions were conducted in Refs.
\cite{shabaev:02:pra,yan:02:jpb,glazov:04:pra,glazov:06:pla}.
Apart from the one-electron contributions to the $g$ factor of $2s$ state,
the effects of the interelectronic interaction should be taken into account in three-electron ions.
The one-photon exchange correction was evaluated in the framework of QED in Ref. \cite{shabaev:02:pra}.
The higher-order contributions of the interelectronic interaction were calculated
by means of the large-scale configuration-interaction Dirac-Fock-Sturm method in Ref. \cite{glazov:04:pra}.
Still, the uncertainty associated with these contributions amounts to more
than half of the total theoretical uncertainty.
The effect of the interelectronic interaction on the QED corrections was treated
within two approaches. For low-$Z$ ions the perturbation theory to the leading orders
in $\aZ$ was employed \cite{yan:02:jpb,glazov:04:pra}.
For middle-$Z$ and high-$Z$ ions more accurate results were obtained by evaluating
the one-electron QED corrections in an effective screening potential \cite{glazov:06:pla}.
Nevertheless, for all values of $Z$ the uncertainty of the screened QED effects
contributes significantly to the total uncertainty, and the rigorous evaluation
of these effects is in demand.

Hyperfine structure of highly charged ions comprises another sensitive tool
for probing QED effects in strong fields.
Accurate measurements of the ground-state hyperfine splittings were performed
for several H-like ions, including $^{209}$Bi, $^{165}$Ho, $^{185}$Re,
$^{187}$Re, $^{207}$Pb, $^{203}$Tl, and $^{205}$Tl in Refs.
\cite{klaft:94:prl,crespo:96:prl,crespo:98:pra,seelig:98:prl,beiersdorfer:01:pra}.
These experiments motivated corresponding theoretical investigations
\cite{persson:96:prl,blundell:97:prl,shabaev:97:pra,sunnergren:98:pra,boucard:00:epjd,
shabaev:00:hi,sapirstein:01:pra,shabaev:01:prl}.
The theoretical uncertainty of the hyperfine splitting is also dominated
by the nuclear effects, mainly by the Bohr-Weisskopf effect \cite{shabaev:97:pra}.
The theoretical investigations have shown \cite{shabaev:01:prl} that simultaneous
studies of the hyperfine splitting in H-like and Li-like ions of the same isotope
can significantly improve the accuracy of the theoretical prediction.
As a result, the ground-state hyperfine splitting in Li-like bismuth was predicted
to a high accuracy using the experimental result for the $1s$ hyperfine splitting
in H-like bismuth \cite{shabaev:00:hi,sapirstein:01:pra,shabaev:01:prl}.
The indirect measurement of the hyperfine splitting of lithiumlike bismuth
performed in Livermore yielded value of 820(26) meV \cite{beiersdorfer:98:prl}.
Determination of this splitting to a much higher accuracy (of about $10^{-7}$)
is planned at GSI in the framework of the HITRAP project \cite{winters:07:cjp}.
This requires further improvements of the theoretical predictions for Li-like ions
and, in particular, evaluations of the QED screening effect. An approximate treatment
of this effect was accomplished in Refs. \cite{sapirstein:01:pra,
sapirstein:08:pra,oreshkina:07:os,kozhedub:07:pra,oreshkina:08:pla,volotka:08:pra}
by employing an effective screening potential in calculations of the one-loop
self-energy and vacuum-polarization diagrams.

Rigorous evaluation of the two-electron self-energy and vacuum-polarization diagrams
(Figs. \ref{fig:two-el-se} and \ref{fig:two-el-vp}) remained a challenge for theory
until recently. In our Letter \cite{volotka:09:prl} the complete $\aZ$-dependent
contributions of the two-electron self-energy diagrams and a dominant part
of the two-electron vacuum-polarization diagrams have been evaluated
for the hyperfine structure of Li-like bismuth and for the $g$ factor of Li-like lead.
In the present paper we describe in detail the evaluation of the screened
quantum electrodynamical correction in presence of external magnetic field.
Furthermore, we extend our calculations to the wide range of nuclear charge
$Z=20$ -- $83$ in case of the hyperfine splitting. The accuracy of the theoretical
prediction for the specific difference of the hyperfine splittings of H-
and Li-like bismuth is improved. As to the $g$ factor, we present
the results for lead and uranium ions.

The relativistic units ($\hbar=1, c=1, m=1$) and the Heaviside charge unit
[$\alpha=e^2/(4\pi), e<0$] are used throughout the paper.

%
\section{Formulation}
\label{sec:basics}
%

A systematic derivation of the QED corrections in a fully
relativistic approach requires the use of perturbation theory
starting with a one-electron approximation,
described by the Dirac equation,
\be
\label{eq:dirac}
  \left[-i\balpha\cdot\bnabla + \beta m + V(\bfx) \right] \psi_n(\bfx)
  = \veps_n \psi_n(\bfx)
\,.
\ee
In our present treatment the binding potential $V(\bfx)=V(|\bfx|)$
denotes the nuclear potential only.
The interaction of the electrons with the quantized electromagnetic
field and the interelectronic-interaction effects are accounted for
by the perturbation theory. In this way we obtain quantum electrodynamics
in the Furry picture.
To derive the formal expressions for the perturbation theory terms,
we employ the two-time Green-function method \cite{shabaev:02:prep}.

\begin{figure}
\includegraphics{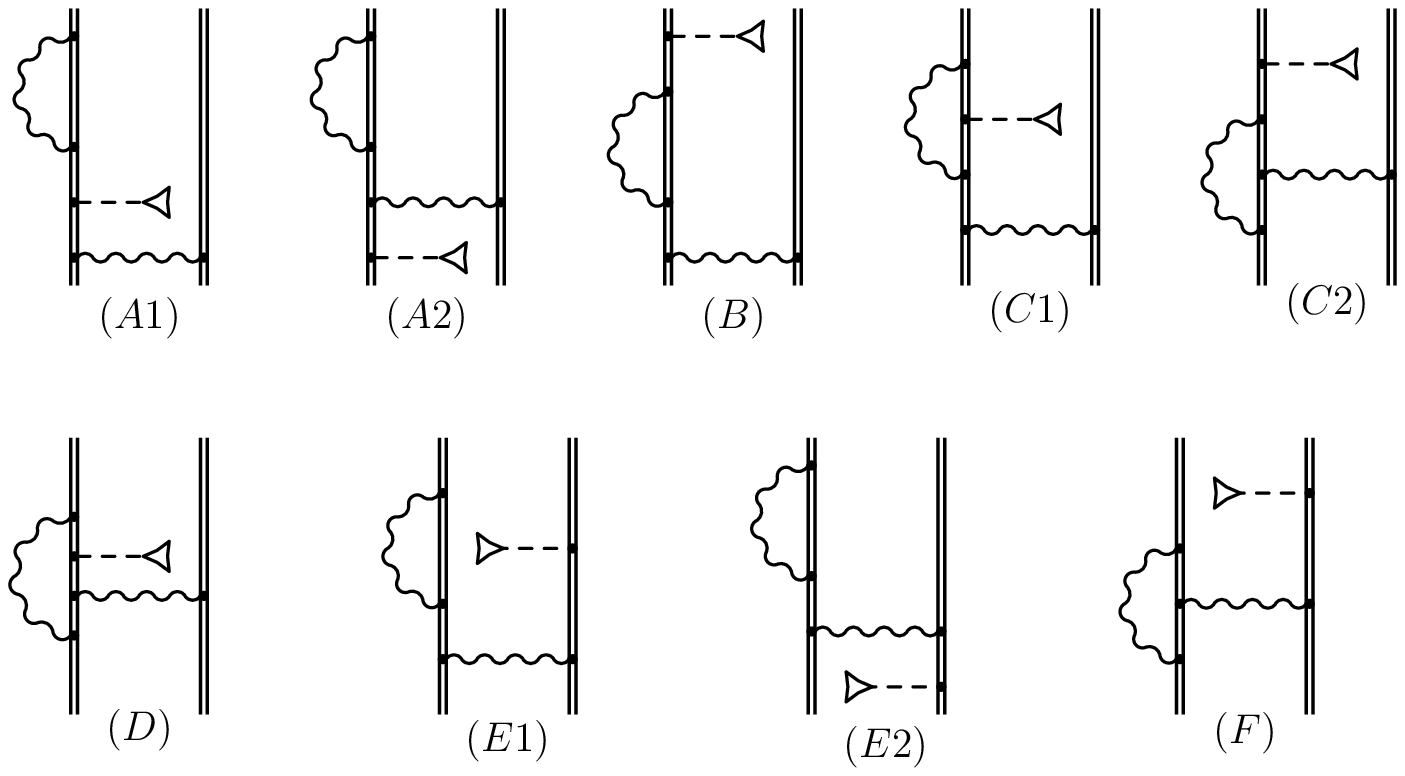}
\caption {Feynman diagrams representing the screened self-energy
correction in the presence of an external magnetic field.
The wavy line indicates the photon propagator and the double line
indicates the electron propagators in the Coulomb field.
The dashed line terminated with the triangle denotes
the interaction with the magnetic field.}
\label{fig:two-el-se}
\end{figure}

\begin{figure}
\includegraphics{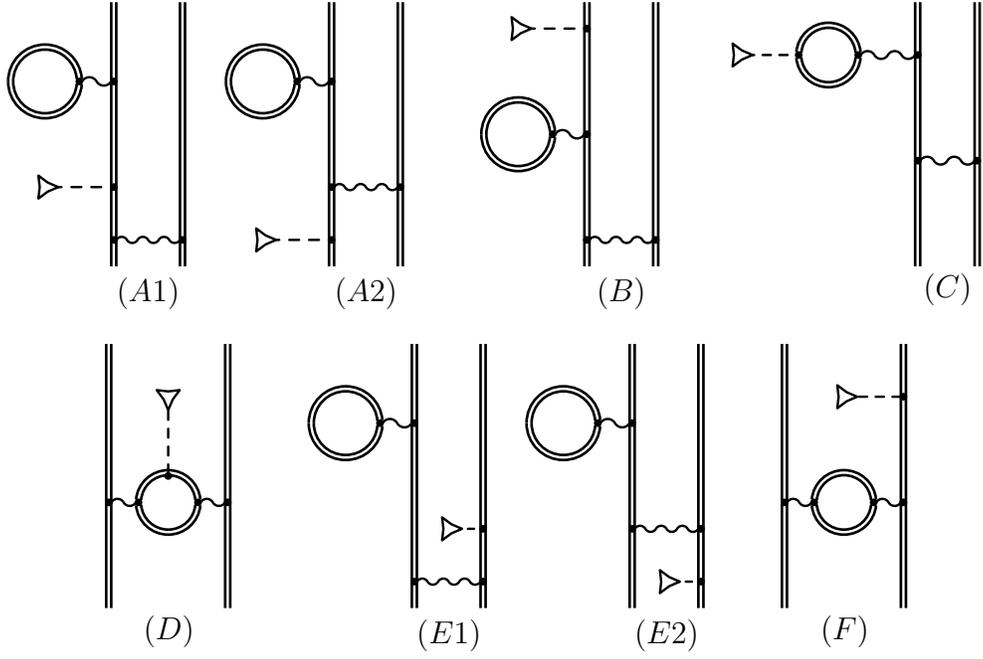}
\caption {Feynman diagrams representing the screened vacuum-polarization
correction in the presence of an external magnetic field.
The notations are the same as in Fig. 1.
}
\label{fig:two-el-vp}
\end{figure}

\begin{figure}
\includegraphics[scale=0.9]{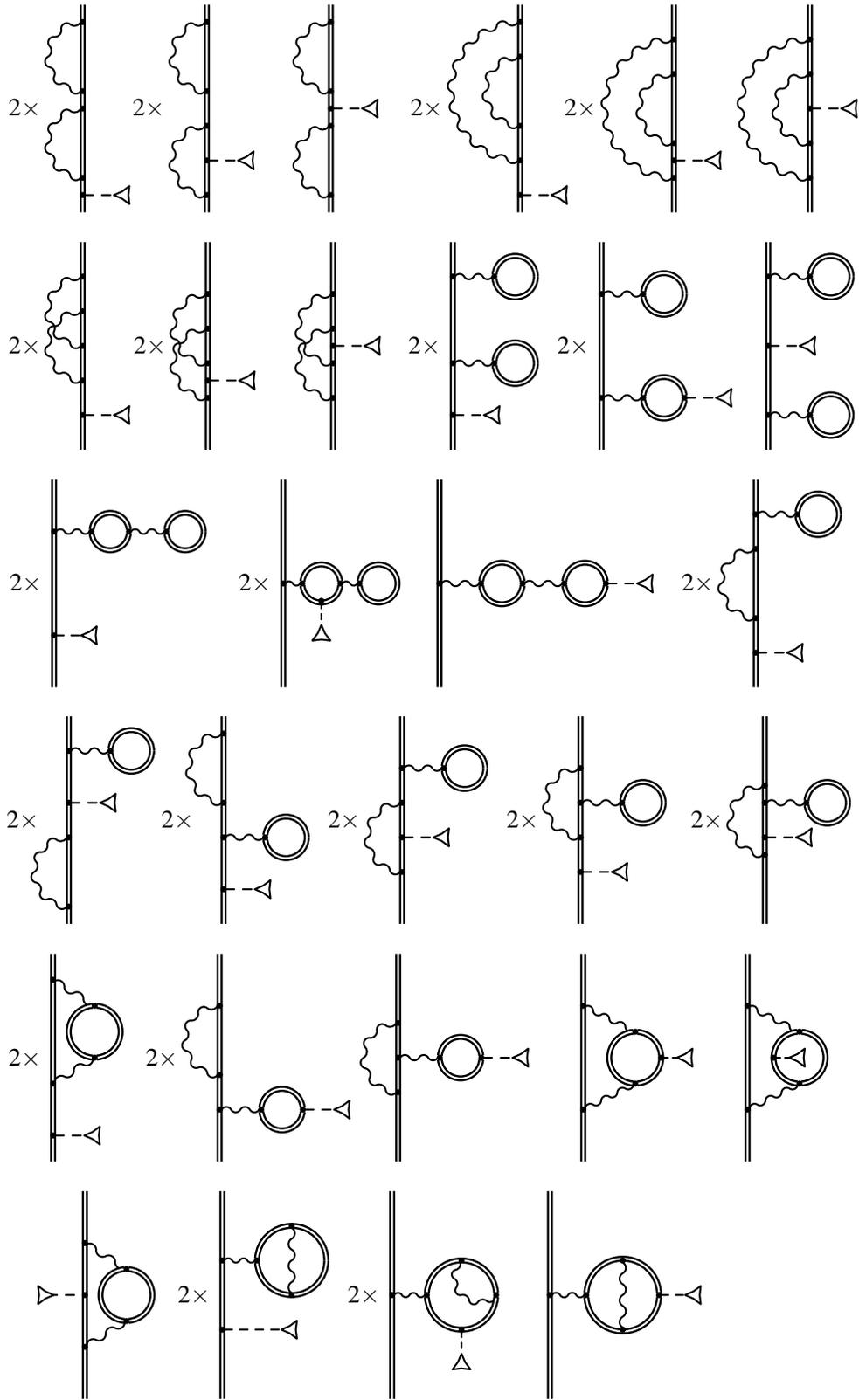}
\caption {Feynman diagrams representing the two-loop QED
corrections in the presence of an external potential.
The notations are the same as in Fig. 1.
}
\label{fig:two-loop}
\end{figure}

The diagrams, which contribute to the screened self-energy and vacuum-polarization
corrections of the first order in $\alpha$ and $1/Z$ in presence of external
magnetic field are depicted in Figs. \ref{fig:two-el-se} and \ref{fig:two-el-vp}.
In order to simplify the derivation we specify the formalism, where the electrons
of the closed shell are regarded as belonging to a new redefined vacuum.
The redefinition of the vacuum results in replacing $i0$ by $-i0$
in the electron propagator denominators corresponding to the closed shell.
In this formalism the one-electron radiative corrections are
incorporated together with the interelectronic-interaction contributions.
In particular, the one-loop two-electron contributions are merged with
the two-loop one-electron contributions. The corresponding diagrams
are depicted in Fig. \ref{fig:two-loop}.
Below we briefly describe the scheme of derivation of the formulas,
corresponding to this set of diagrams within the two-time Green-function method.
In order to obtain the two-electron corrections one may simply consider
the related expressions with the standard definition of the vacuum
and then consequently make a replacement
\be
\label{eq:S-repl}
  \sum_n \frac{\ket{n}\bra{n}}{\veps-u\veps_n} \to
  2\pi i \delta(\veps-\veps_c)\sum_c \ket{c}\bra{c}
\ee
for each of the electron propagators inside the loops.
Here and in the following the notation $u=1-i0$ is used.
The summation over $c$ is performed over all electrons of the closed shell.

To zeroth-order approximation the state $\ket{a}$ of the electron is defined
by the Dirac equation (\ref{eq:dirac}). The energy shift of an isolated level
due to the interaction is given by \cite{shabaev:02:prep}
\be
\label{eq:dE}
  \Delta E_a = \frac{(2\pi i)^{-1}\intG\rmd E (E-E_a^{(0)}) \Delta g_{aa}(E)}
    {1+(2\pi i)^{-1}\intG\rmd E \Delta g_{aa}(E)}
\,.
\ee
In our case the unperturbed energy $E_a^{(0)}$  is the Dirac energy $\veps_a$
from Eq. (\ref{eq:dirac}).
The contour $\Gamma$ surrounds only the pole $E=E_a^{(0)}$,
$\Delta g_{aa}(E) = g_{aa}(E) - g^{(0)}_{aa}(E)$,
$g_{aa}(E) = \matrixel{\psi_a}{g(E)}{\psi_a}$, and $\psi_a$ is the unperturbed wave function.
The time-Fourier transform of the Green function is defined as
\be
  g(E,\bfxpr,\bfx) \delta(E-\Epr) = \frac{1}{2\pi i} \intinff \rmd t \, \rmd \tpr \,
    \exp(i\Epr\tpr-iEt) \matrixel{0}{T\psi(\tpr,\bfxpr)\cpsi(t,\bfx)}{0}
\,.
\ee
The Feynman rules for the Green function are given in Ref. \cite{shabaev:02:prep}.
The energy shift $\Delta E_a$ and the Green function $g(E)$ are to be expanded
as a power series in $\alpha$,
\be
  \Delta E_a = \Delta E_a^{(1)} + \Delta E_a^{(2)} + \Delta E_a^{(3)} + \dots
\,,
\\
  \Delta g_{aa}(E) = \Delta g_{aa}^{(1)}(E) + \Delta g_{aa}^{(2)}(E) + \Delta g_{aa}^{(3)}(E) + \dots
\,.
\ee
Then from Eq. (\ref{eq:dE}) we find the first-, second- and third-order terms
of the energy shift,
\be
\label{eq:dE-1}
  \Delta E^{(1)} &=& \frac{1}{2\pi i}\intG\rmd E (E-E_a^{(0)}) \Delta g^{(1)}_{aa}(E)
\,,
\\
\label{eq:dE-2}
  \Delta E^{(2)} &=& \frac{1}{2\pi i}\intG\rmd E (E-E_a^{(0)}) \Delta g^{(2)}_{aa}(E)
\nonumber\\
&-& \frac{1}{2\pi i}\intG\rmd E (E-E_a^{(0)}) \Delta g^{(1)}_{aa}(E)\,
    \frac{1}{2\pi i}\intG\rmd \Epr \Delta g^{(1)}_{aa}(\Epr)
\,,
\\
\label{eq:dE-3}
  \Delta E^{(3)} &=& \frac{1}{2\pi i}\intG\rmd E (E-E_a^{(0)}) \Delta g^{(3)}_{aa}(E)
\nonumber\\
&-& \frac{1}{2\pi i}\intG\rmd E (E-E_a^{(0)}) \Delta g^{(2)}_{aa}(E)\,
    \frac{1}{2\pi i}\intG\rmd \Epr \Delta g^{(1)}_{aa}(\Epr)
\nonumber\\
&-& \frac{1}{2\pi i}\intG\rmd E (E-E_a^{(0)}) \Delta g^{(1)}_{aa}(E)
\nonumber\\
&\times&
    \left[ \frac{1}{2\pi i}\intG\rmd \Epr \Delta g^{(2)}_{aa}(\Epr)
         - \left( \frac{1}{2\pi i}\intG\rmd \Epr \Delta g^{(1)}_{aa}(\Epr) \right)^2
    \right]
\,.
\ee
Equation (\ref{eq:dE-3}) and the Feynman rules for $g(E)$ yield the formal expressions
for the total contribution of the two-loop diagrams presented in Fig. \ref{fig:two-loop}.
Consequently replacing each electron propagator according to Eq. (\ref{eq:S-repl})
we obtain the formal expressions for the two-electron one-loop diagrams
displayed in Figs. \ref{fig:two-el-se} and \ref{fig:two-el-vp}.

For brevity we introduce the operator
\be
  I(\omega) = e^2 \alpha^{\mu} \alpha^{\nu} D_{\mu\nu} (\omega)
\,,
\ee
where $\alpha^{\mu}=(1,\balpha)$ are the Dirac matrices,
and $D_{\mu\nu}$ is the photon propagator.
It is given by
\be
  D_{\mu\nu}(\omega,\bfx_{12}) = g_{\mu\nu}
    \frac{\exp(i\tomega|\bfx_{12}|)}{4\pi|\bfx_{12}|}
\,,
\ee
in the Feynman gauge and by
\be
&&  D_{00}(\omega,\bfx_{12}) = \frac{1}{4\pi|\bfx_{12}|}
\,,\qquad
    D_{i0}(\omega,\bfx_{12}) = D_{0k}(\omega,\bfx_{12}) = 0
\,,
\nonumber\\
&&  D_{ik}(\omega,\bfx_{12}) = \delta_{ik}
    \frac{\exp(i\tomega|\bfx_{12}|)}{4\pi|\bfx_{12}|}
  + \nabla_{1\,i} \nabla_{2\,k}
    \frac{1-\exp(i\tomega|\bfx_{12}|)}{4\pi\omega^2|\bfx_{12}|}
\,,
\qquad
  (i,k=1,2,3)
\,,
\ee
in the Coulomb gauge.
Here $\bfx_{12}=\bfx_1-\bfx_2$, $\tomega=\sqrt{\omega^2+i0}$.
The branch of the square root is fixed by the condition
${\rm Im}\,\tomega>0$.
In order to handle the infrared divergencies it is convenient
to introduce the photon mass $\mu$. In the Feynman gauge, it results
in the replacement $\sqrt{\omega^2+i0}\to\sqrt{\omega^2-\mu^2+i0}$.
The limit $\mu \to 0$ should be taken after removing the divergencies.
The operator $I(\omega)$ has the following symmetry properties:
\be
&&  I(\omega) = I(-\omega)
\,,\nonumber\\
&&  I'(\omega) \equiv \ddo I(\omega) = -I'(-\omega)
\,,\nonumber\\
&&  I''(\omega) \equiv \dddo I(\omega) = I''(-\omega)
\,.
\ee
The interaction with the external magnetic field can be represented by an operator $T_0$.
For both cases under consideration ($g$ factor and hyperfine splitting) it is
proportional to $\left[\bfr\times\balpha\right]_z$, what defines the angular momentum
structure. Explicit formulas for $T_0$ will be given in Sec. \ref{sec:results}.
%
%
\section{Screened self-energy}
\label{sec:scr-se}
%

The diagrams of the screened self-energy correction, corresponding to the first
term in Eq. (\ref{eq:dE-3}), are shown in Fig. \ref{fig:two-el-se}.
We divide the total contribution of these diagrams
into the reducible and irreducible parts. The irreducible part is the sum
of the terms where the energies of the intermediate states are different from
the energy of the initial state. The reducible part is the remainder.
We denote the irreducible parts of each diagram $A$ -- $F$ by the same letter:
$\dEScrSEX{A-F}$. It is convenient to divide them into three groups, according
to the number of the $\omega$-dependent denominators ($\omega$ is the virtual
photon energy, over which the integration is performed). The terms with only
one denominator ($A$, $B$ and $E$) are referred to as a "modified self-energy" terms,
since all of them have the form of a matrix element of the self-energy operator,
$\matrixel{X}{\Sigma(\veps)}{Y}$. The terms with two denominators ($C$ and $F$)
are denoted as a "modified vertex", and the diagram $D$ with three denominators
is denoted as a "double-vertex". The reducible parts are considered together
with the related contributions that arise from the second and the third terms
in Eq. (\ref{eq:dE-3}). Their sum is divided into three parts: $G$, $H$, and $I$,
according to the number of the $\omega$-dependent denominators. These three parts
are considered together with the "modified self-energy", "modified vertex",
and "double-vertex" terms, respectively.
Finally, the total contribution of the two-electron self-energy diagrams is given by
\be
\label{eq:dEScrSE}
  \dEScrSE = \sum_{b} \sum_{PQ} (-1)^{P+Q}
    \left( \dEA + \dEB + \dEC + \dED
\right.
\nonumber\\
\left.
  + \dEE + \dEF + \dEG + \dEH + \dEI
    \right)
\,.
\ee
Here $P$ and $Q$ are the permutation operators, interchanging the valence ($a$)
and the core ($b$) electron states, $(-1)^P$ is the sign of the permutation $P$.
The summation over $b$ runs over two core electron states with different projections
of the total angular momentum. In what follows, we will also use the notation
$\DeltaQbPb\equiv\veps_{Qb}-\veps_{Pb}$.

One has to pay special attention to the ultraviolet (UV) divergencies, that arise
in the formal expressions under consideration.
First, we introduce the unrenormalized self-energy operator,
\be
\label{eq:sigma}
  \matrixel{p}{\Sigma(\veps)}{q} =
    \frac{i}{2\pi} \intinf \rmd\omega \sum_{n}
    \frac{\matrixel{p n}{I(\omega)}{n q}}{\veps-\omega-u\veps_{n}}
\,.
\ee
Here and below the integration over $\omega$ is carried out
from $-\infty$ to $+\infty$.
Every diagram involving the self-energy loop has to be considered together with
the corresponding diagram with the mass counterterm, what results in the replacement
$\Sigma(\veps) \to \Sigma_{\rm R}(\veps) = \Sigma(\veps) - \gamma^0 \delta m$.
The matrix elements of $\Sigma_{\rm R}(\veps)$ still have the divergent part:
\be
\label{eq:sigma-uv1}
  \matrixel{p}{\Sigma_{\rm R}(\veps)}{q} =
    B^{(1)} \matrixel{p}{\left[\veps - \balpha\cdot\bfp - \beta m - V(\bfx)\right]}{q}
    + \mbox{finite part}
\,,
\ee
where $B^{(1)}$ is the UV-divergent constant.
Assuming that $\ket{p}$ and $\ket{q}$ obey the Dirac equation (\ref{eq:dirac}) we have,
\be
\label{eq:sigma-uv2}
  \matrixel{p}{\Sigma_{\rm R}(\veps)}{q} =
    B^{(1)} (\veps - \veps_p) \delta_{pq}
    + \mbox{finite part}
\,.
\ee
In order to isolate the divergent part we follow the potential-expansion approach \cite{snyderman:91:ann}.
The finite part of the self-energy matrix element is then divided into the zero-
and one-potential terms evaluated in momentum space, and the many-potential term
evaluated in coordinate space.
The diagrams with one vertex inside the self-energy loop also suffer from UV-divergencies.
It can be shown that for an arbitrary operator $U$,
\be
\label{eq:gamma-uv}
  \frac{i}{2\pi} \intinf \rmd \omega \sum_{n_{1,2}}
    \frac{\matrixel{p n_2}{I(\omega)}{n_1 q}\matrixel{n_1}{U}{n_2}}
         {(\veps-\omega-u\vepsn{1})(\veps-\omega-u\vepsn{2})}
  = L^{(1)} \matrixel{p}{U}{q}
    + \mbox{finite part}
\,.
\ee
In our case $U$ is either $T_0$ or $I(\Delta)$
(in the latter case $\matrixel{p}{U}{q} \equiv \matrixel{p r}{I(\Delta)}{q s}$).
Due to the Ward identity we have $L^{(1)} = - B^{(1)}$.
The finite part of the vertex contributions is divided into the zero-potential term
evaluated in momentum space and the many-potential term evaluated in coordinate space.

From Eq. (\ref{eq:sigma-uv2}) one can see that for the first-order self-energy
correction the divergent part is zero.
However, in case of the higher-order diagrams under consideration the contributions
of particular diagrams are divergent. Nevertheless, as it is shown below,
the sum of all the contributions to the screened self-energy correction is finite.

\subsection{"Modified self-energy" diagrams}

The irreducible parts of the diagrams $A$, $B$ and $E$ can be presented
as the matrix elements of the self-energy operator $\matrixel{X}{\Sigma(\veps)}{Y}$
with various wave functions $\bra{X}$ and $\ket{Y}$. The formulas for these
contributions are as follows,
\be
\label{eq:a}
  \dEA = \dEAi{1} + \dEAi{2}
\,,
\ee
\be
\label{eq:a1}
  \dEAi{1} = 2 \,{\sum_{n_{1,2}}}'
    \matrixel{Pa}{\Sigma(\veps_{Pa})}{n_1}
    \frac{\matrixel{n_1}{T_0}{n_2}\matrixel{n_2 Pb}{I(\DeltaQbPb)}{Qa Qb}}
         {(\veps_{Pa}-\vepsn{1})(\veps_{Pa}-\vepsn{2})}
\,,
\ee
\be
\label{eq:a2}
  \dEAi{2} = 2 \,{\sum_{n_{1,2}}}'
    \matrixel{Pa}{\Sigma(\veps_{Pa})}{n_1}
    \frac{\matrixel{n_1 Pb}{I(\DeltaQbPb)}{n_2 Qb}\matrixel{n_2}{T_0}{Qa}}
         {(\veps_{Pa}-\vepsn{1})(\veps_{Qa}-\vepsn{2})}
\,,
\ee
\be
\label{eq:b}
  \dEB = 2 \,{\sum_{n_{1,2}}}'
    \frac{\matrixel{Pa}{T_0}{n_1}}{\veps_{Pa}-\vepsn{1}}
    \matrixel{n_1}{\Sigma(\veps_{Pa})}{n_2}
    \frac{\matrixel{n_2 Pb}{I(\DeltaQbPb)}{Qa Qb}}{\veps_{Pa}-\vepsn{2}}
\,,
\ee
\be
\label{eq:e}
  \dEE = \dEEi{1} + \dEEi{2}
\,,
\ee
\be
\label{eq:e1}
  \dEEi{1} = 2 \,{\sum_{n_{1,2}}}'
    \matrixel{Pa}{\Sigma(\veps_{Pa})}{n_1}
    \frac{\matrixel{Pb}{T_0}{n_2} \matrixel{n_1 n_2}{I(\DeltaQbPb)}{Qa Qb}}
         {(\veps_{Pa}-\vepsn{1})(\veps_{Pb}-\vepsn{2})}
\,,
\ee
\be
\label{eq:e2}
  \dEEi{2} = 2 \,{\sum_{n_{1,2}}}'
    \matrixel{Pa}{\Sigma(\veps_{Pa})}{n_1}
    \frac{\matrixel{n_1 Pb}{I(\DeltaQbPb)}{Qa n_2} \matrixel{n_2}{T_0}{Qb}}
         {(\veps_{Pa}-\vepsn{1})(\veps_{Qb}-\vepsn{2})}
\,.
\ee
All the reducible terms of the similar structure are denoted as $\dEG$,
\be
\label{eq:g}
  \dEG = \dEGi{1} + \dEGi{2} + \dEGi{3}
\,,
\ee
\be
\label{eq:g1}
  \dEGi{1} = - 2 \, {\sum_{n_1}}'
    \frac{1}{(\veps_{Pa}-\vepsn{1})^2}
    \Bigl[
    \matrixel{Pa}{\Sigma(\veps_{Pa})}{n_1}
    \matrixel{n_1}{T_0}{Pa}
    \matrixel{Pa Pb}{I(\DeltaQbPb)}{Qa Qb}
\nonumber\\
  + \matrixel{Pa}{\Sigma(\veps_{Pa})}{n_1}
    \matrixel{n_1 Pb}{I(\DeltaQbPb)}{Qa Qb}
    \matrixel{Pa}{T_0}{Pa}
\nonumber\\
  + \matrixel{Pa}{\Sigma(\veps_{Pa})}{Pa}
    \matrixel{Pa}{T_0}{n_1}
    \matrixel{n_1 Pb}{I(\DeltaQbPb)}{Qa Qb}
    \Bigr]
\,,
\ee
\be
\label{eq:g2}
  \dEGi{2} = 2 \, {\sum_{n_1}}'
    \frac{1}{\veps_{Pa}-\vepsn{1}}
  \Bigl[
    \matrixel{Pa}{\Sigma(\veps_{Pa})}{n_1}
    \matrixel{n_1}{T_0}{Pa}
    \matrixel{Pa Pb}{I'(\DeltaQbPb)}{Qa Qb}
\nonumber\\
  + \matrixel{Pa}{\Sigma(\veps_{Pa})}{n_1}
    \matrixel{n_1 Pb}{I'(\DeltaQbPb)}{Qa Qb}
    \Bigl( \matrixel{Qb}{T_0}{Qb} - \matrixel{Pb}{T_0}{Pb} \Bigr)
  \Bigr]
\nonumber\\
  + 2 \,\matrixel{Pa}{\Sigma(\veps_{Pa})}{Pa} \,{\sum_{n_1}}'
      \left[ \frac{\matrixel{Pa Pb}{I'(\DeltaQbPb)}{n_1 Qb}
                   \matrixel{n_1}{T_0}{Qa}}{\veps_{Qa}-\vepsn{1}}
  \right.
\nonumber\\
  \left.
           + \frac{\matrixel{Pa Pb}{I'(\DeltaQbPb)}{Qa n_1}
                   \matrixel{n_1}{T_0}{Qb}}{\veps_{Qb}-\vepsn{1}}
      \right]
\,,
\ee
\be
\label{eq:g3}
  \dEGi{3} = \matrixel{Pa}{\Sigma(\veps_{Pa})}{Pa}
    \matrixel{Pa Pb}{I''(\DeltaQbPb)}{Qa Qb}
    \Bigl( \matrixel{Qb}{T_0}{Qb} - \matrixel{Pb}{T_0}{Pb} \Bigr)
\,.
\ee
Equations (\ref{eq:a})--(\ref{eq:g3}) possess ultraviolet (UV) divergences.
Taking into account the mass counterterm and employing Eq. (\ref{eq:sigma-uv2})
we find that $\dEB$ has a non-zero UV-divergent part,
\be
\label{eq:b-uv}
  \dEB({\rm UV}) = 2\,B^{(1)}\,{\sum_{n_{1}}}'
    \frac{\matrixel{Pa}{T_0}{n_1}\matrixel{n_1 Pb}{I(\DeltaQbPb)}{Qa Qb}}
         {\veps_{Pa}-\vepsn{1}}
\,.
\ee
By the end of the next subsection we will show that the sum of all the UV-divergent terms is zero.

\subsection{"Modified vertex" diagrams}

For the irreducible parts of the diagrams $C$ and $F$ we have,
\be
\label{eq:c}
  \dEC = \dECi{1} + \dECi{2}
\,,
\ee
\be
\label{eq:c1}
  \dECi{1} = 2 \,\frac{i}{2\pi} \intinf \rmd \omega
    \sum_{n_{1,2,3}}^{\vepsn{3}\neq\veps_{Pa}}
    \frac{\matrixel{Pa n_2}{I(\omega)}{n_1 n_3}\matrixel{n_1}{T_0}{n_2} }
         {(\veps_{Pa}-\omega-u\vepsn{1})(\veps_{Pa}-\omega-u\vepsn{2})}
    \frac{\matrixel{n_3 Pb}{I(\DeltaQbPb)}{Qa Qb}}{(\veps_{Pa}-\vepsn{3})}
\,,
\ee
\be
\label{eq:c2}
  \dECi{2} = 2 \,\frac{i}{2\pi} \intinf \rmd \omega
    \sum_{n_{1,2,3}}^{\vepsn{3}\neq\veps_{Qa}}
    \frac{\matrixel{Pa n_2}{I(\omega)}{n_1 n_3}\matrixel{n_1 Pb}{I(\DeltaQbPb)}{n_2 Qb} }
         {(\veps_{Pa}-\omega-u\vepsn{1})(\veps_{Qa}-\omega-u\vepsn{2})}
    \frac{\matrixel{n_3}{T_0}{Qa}}{(\veps_{Qa}-\vepsn{3})}
\,,
\ee
\be
\label{eq:f}
  \dEF = 2 \,\frac{i}{2\pi} \intinf \rmd \omega
    \sum_{n_{1,2,3}}^{\vepsn{3}\neq\veps_{Qb}}
    \frac{\matrixel{Pa n_2}{I(\omega)}{n_1 Qa}\matrixel{n_1 Pb}{I(\DeltaQbPb)}{n_2 n_3} }
         {(\veps_{Pa}-\omega-u\vepsn{1})(\veps_{Qa}-\omega-u\vepsn{2})}
    \frac{\matrixel{n_3}{T_0}{Qb}}{(\veps_{Qb}-\vepsn{3})}
\,.
\ee
Since these diagrams have one vertex inside the self-energy loop,
the corresponding expressions have the following structure
of the $\omega$-dependent denominators:
$(\Delta_1 - \omega)^{-1}(\Delta_2 - \omega)^{-1}$.
All the reducible terms that have similar structure are denoted as $\dEH$,
\be
\label{eq:h}
  \dEH = \dEHi{1} + \dEHi{2} + \dEHi{3}
\,,
\ee
\be
\label{eq:h1}
  \dEHi{1} = \frac{i}{2\pi} \intinf \rmd \omega
    \sum_{n_{1,2}}
    \frac{\matrixel{Pa n_2}{I(\omega)}{n_1 Pa}\matrixel{n_1}{T_0}{n_2}}
         {(\veps_{Pa}-\omega-u\vepsn{1})(\veps_{Pa}-\omega-u\vepsn{2})}
    \matrixel{Pa Pb}{I'(\DeltaQbPb)}{Qa Qb}
\,,
\ee
\be
\label{eq:h2}
  \dEHi{2} = \frac{i}{2\pi} \intinf \rmd \omega
    \sum_{n_{1,2}}
    \frac{\matrixel{Pa n_2}{I(\omega)}{n_1 Qa}\matrixel{n_1 Pb}{I'(\DeltaQbPb)}{n_2 Qb}}
         {(\veps_{Pa}-\omega-u\vepsn{1})(\veps_{Qa}-\omega-u\vepsn{2})}
\nonumber\\
    \times\Bigl( \matrixel{Qb}{T_0}{Qb} - \matrixel{Pb}{T_0}{Pb} \Bigl)
\,,
\ee
\be
\label{eq:h3}
  \dEHi{3} = 2 \, {\sum_{n_1}}' \frac{1}{\veps_{Pa}-\vepsn{1}}
    \Bigl[
    \matrixel{Pa}{\Sigma'(\veps_{Pa})}{n_1}
    \matrixel{n_1}{T_0}{Pa}
    \matrixel{Pa Pb}{I(\DeltaQbPb)}{Qa Qb}
\nonumber\\
  + \matrixel{Pa}{\Sigma'(\veps_{Pa})}{n_1}
    \matrixel{n_1 Pb}{I(\DeltaQbPb)}{Qa Qb}
    \matrixel{Pa}{T_0}{Pa}
    \Bigr]
\nonumber\\
  + \matrixel{Pa}{\Sigma'(\veps_{Pa})}{Pa} \left\{
    2\,{\sum_{n_1}}'
    \left[ \frac{\matrixel{Pa}{T_0}{n_1}\matrixel{n_1 Pb}{I(\DeltaQbPb)}{Qa Qb}}
                {\veps_{Pa}-\vepsn{1}}
\right.
\right.
\nonumber\\
\left.
         + \frac{\matrixel{Pb}{T_0}{n_1}\matrixel{Pa n_1}{I(\DeltaQbPb)}{Qa Qb}}
                {\veps_{Pb}-\vepsn{1}}
    \right]
\nonumber\\
\left.
  + \matrixel{Pa Pb}{I'(\DeltaQbPb)}{Qa Qb}
    \Bigl( \matrixel{Pa}{T_0}{Pa} + \matrixel{Qb}{T_0}{Qb} - \matrixel{Pb}{T_0}{Pb} \Bigl)
\vphantom{{\sum_{n_1}}'}
\right\}
\,.
\ee
Equations (\ref{eq:c})--(\ref{eq:h3}) diverge both in ultraviolet and infrared regions.
The UV-divergent terms are:
\be
\label{eq:c1-uv}
  \dECi{1}({\rm UV}) = 2\,L^{(1)}\,{\sum_{n_{1}}}'
    \frac{\matrixel{Pa}{T_0}{n_1}\matrixel{n_1 Pb}{I(\DeltaQbPb)}{Qa Qb}}{\veps_{Pa}-\vepsn{1}}
\,,
\ee
\be
\label{eq:c2-uv}
  \dECi{2}({\rm UV}) = 2\,L^{(1)}\,{\sum_{n_{1}}}'
    \frac{\matrixel{Pa Pb}{I(\DeltaQbPb)}{n_1 Qb}\matrixel{n_1}{T_0}{Qa}}{\veps_{Qa}-\vepsn{1}}
\,,
\ee
\be
\label{eq:f-uv}
  \dEF({\rm UV}) = 2\,L^{(1)}\,{\sum_{n_{1}}}'
    \frac{\matrixel{Pa Pb}{I(\DeltaQbPb)}{Qa n_1}\matrixel{n_1}{T_0}{Qb}}{\veps_{Qb}-\vepsn{1}}
\,,
\ee
\be
\label{eq:h1-uv}
  \dEHi{1}({\rm UV}) = L^{(1)}\,\matrixel{Pa}{T_0}{Pa}\matrixel{Pa Pb}{I'(\DeltaQbPb)}{Qa Qb}
\,,
\ee
\be
\label{eq:h2-uv}
  \dEHi{2}({\rm UV}) = L^{(1)}\,\matrixel{Pa Pb}{I'(\DeltaQbPb)}{Qa Qb}
    \Bigl( \matrixel{Qb}{T_0}{Qb} - \matrixel{Pb}{T_0}{Pb} \Bigl)
\,,
\ee
\be
\label{eq:h3-uv}
  \dEHi{3}({\rm UV}) = 2\,B^{(1)}\,{\sum_{n_{1}}}'
    \left( \frac{\matrixel{Pa}{T_0}{n_1}\matrixel{n_1 Pb}{I(\DeltaQbPb)}{Qa Qb}}
                {\veps_{Pa}-\vepsn{1}}
\right.
\nonumber\\
\left.
         + \frac{\matrixel{Pb}{T_0}{n_1}\matrixel{Pa n_1}{I(\DeltaQbPb)}{Qa Qb}}
                {\veps_{Pb}-\vepsn{1}}
    \right)
\nonumber\\
  + B^{(1)}\,\matrixel{Pa Pb}{I'(\DeltaQbPb)}{Qa Qb}
    \Bigl( \matrixel{Pa}{T_0}{Pa} + \matrixel{Qb}{T_0}{Qb} - \matrixel{Pb}{T_0}{Pb} \Bigr)
\,.
\ee
One can see that the sum of the UV-divergent terms (\ref{eq:b-uv}), (\ref{eq:c1-uv})--(\ref{eq:h3-uv})
is zero.

There are also infrared (IR) divergencies in Eqs. (\ref{eq:c})--(\ref{eq:h3}).
They arise when the energies of the intermediate states $n$ are equal to the energies
of the reference states $a$ or $b$, what leads to the factor ${1}/{(\omega-i0)^2}$.
In order to handle these divergencies we introduce a non-zero photon mass $\mu$
and isolate analytically the terms proportional to $\ln\mu$. Similar terms arise
in Eqs. (\ref{eq:d})--(\ref{eq:i}) and should be considered together to yield a finite result.

\subsection{"Double-vertex" diagrams}

Finally, we consider the diagram $D$ with two vertices inside the self-energy loop,
\be
\label{eq:d}
  \dED = 2 \frac{i}{2\pi} \intinf \rmd \omega
    \sum_{n_{1,2,3}}
    \frac{\matrixel{Pa n_3}{I(\omega)}{n_1 Qa}\matrixel{n_1 Pb}{I(\DeltaQbPb)}{n_2 Qb}\matrixel{n_2}{T_0}{n_3} }
         {(\veps_{Pa}-\omega-u\vepsn{1})(\veps_{Qa}-\omega-u\vepsn{2})(\veps_{Qa}-\omega-u\vepsn{3})}
\,.
\ee
The reducible contributions with the third power of $\omega$ in the denominators
are denoted as $\dEI$,
\be
\label{eq:i}
  \dEI = \dEIi{1} + \dEIi{2} + \dEIi{3}
\,,
\ee
\be
\label{eq:i1}
  \dEIi{1} = - \frac{i}{2\pi} \intinf \rmd\omega \sum_{n_{1,2}}
    \frac{\matrixel{Pa n_2}{I(\omega)}{n_1 Pa}\matrixel{n_1}{T_0}{n_2}}
         {(\veps_{Pa}-\omega-u\vepsn{1})(\veps_{Pa}-\omega-u\vepsn{2})}
\nonumber\\
  \times
   \left( \frac{1}{\veps_{Pa}-\omega-u\vepsn{1}}
        + \frac{1}{\veps_{Pa}-\omega-u\vepsn{2}} \right)
   \matrixel{Pa Pb}{I(\DeltaQbPb)}{Qa Qb}
\,,
\ee
\be
\label{eq:i2}
  \dEIi{2} = - \frac{i}{2\pi} \intinf \rmd\omega \sum_{n_{1,2}}
    \frac{\matrixel{Pa n_2}{I(\omega)}{n_1 Qa}
          \matrixel{n_1 Pb}{I(\DeltaQbPb)}{n_2 Qb}}
         {(\veps_{Pa}-\omega-u\vepsn{1})(\veps_{Qa}-\omega-u\vepsn{2})}
\nonumber\\
  \times
   \left( \frac{\matrixel{Pa}{T_0}{Pa}}{\veps_{Pa}-\omega-u\vepsn{1}}
        + \frac{\matrixel{Qa}{T_0}{Qa}}{\veps_{Qa}-\omega-u\vepsn{2}} \right)
\,,
\ee
\be
\label{eq:i3}
  \dEIi{3} = \matrixel{Pa}{\Sigma''(\veps_{Pa})}{Pa}
             \matrixel{Pa}{T_0}{Pa}
             \matrixel{Pa Pb}{I(\DeltaQbPb)}{Qa Qb}
\,.
\ee
These contributions are UV-finite. However, they contain IR-divergent terms
that should be considered together with similar terms from Eqs. (\ref{eq:c1})--(\ref{eq:h}).
Introducing a non-zero photon mass $\mu$ and isolating the terms, proportional to $\ln\mu$
we find that the sum of all the IR-divergent terms
from the terms $C$, $F$, $H$, $D$ and $I$ is finite.

\subsection{Numerical evaluation}

Evaluation of the two-electron self-energy correction requires momentum-space
calculation of the zero- and one-potential terms of the "modified self-energy" contributions
($\dEScrSEX{A}$, $\dEScrSEX{B}$, $\dEScrSEX{E}$, and $\dEScrSEX{G}$)
and zero-potential terms of the "modified vertex" contributions
($\dEScrSEX{C}$, $\dEScrSEX{F}$, and $\dEScrSEX{H}$).
For the zero- and one-potential terms of the "modified self-energy" contributions
we employ the numerical procedure, developed for the self-energy diagram
\cite{blundell:91:pra,yerokhin:99:pra}.

The "magnetic-vertex" part (with the operator $T_0$ in the vertex)
of the "modified vertex" contributions is somewhat similar to the vertex part
of the one-electron self-energy correction to the $g$ factor or the hyperfine splitting.
Hence, our treatment of its zero-potential term is based on the corresponding
calculational procedures developed in Ref. \cite{yerokhin:04:pra} for the $g$ factor
and in Refs. \cite{blundell:97:pra,sunnergren:98:pra} for the hyperfine splitting.
The angular integration, however, required significant generalization,
in both cases, due to the interelectronic-interaction matrix elements.
The "interaction-vertex" part (with the interelectronic-interaction operator
$I(\Delta)$ in the vertex) resembles the vertex part of the two-electron
self-energy correction to the energy levels. Therefore, we have developed
the numerical algorithm for the corresponding zero-potential term on the basis
of the one presented in Ref. \cite{yerokhin:99:pra:scr-se}.
The main difference is the structure of the angular integrals which is
substantially more complicated, due to the interaction with the external magnetic field.

The many-potential terms of the "modified self-energy" and "modified vertex" parts,
as well as the complete contribution of the "double vertex" part are calculated
in the coordinate space. Angular integration and summation over intermediate
angular momentum projections is carried out in the standard way.
The summation over the complete spectrum of the Dirac equation at fixed
angular quantum numbers $\kappa_{1,2,3}$ is performed using the dual-kinetic-balance (DKB)
approach \cite{shabaev:04:prl} with the basis functions constructed from $B$-splines
\cite{sapirstein:96:jpb}.
The infinite summation over $\kappa_{1,2,3}$ is terminated at $|\kappa|=10$--$15$
and the rest of the sum is evaluated by the least-square inverse-polynomial fitting.
In order to perform the integration over $\omega$ we employ two different
contours, thus performing an additional cross-check.
Both of them involve the Wick rotation.
The first contour is the same as in Ref. \cite{blundell:91:pra}.
The integration is performed along the imaginary axis. Besides, the contributions
from the poles arising from the terms with $\veps_n\le\veps_{a,b}$ must be calculated.
The advantage of this contour is that the calculation is less time-consuming,
in particular, due to the fact that ${\rm Im}[I(\omega)]=0$ when ${\rm Re}[\omega]=0$.
The second contour was proposed in Ref. \cite{yerokhin:99:pra}. The integration is performed
along the line $C_{\rm H}$ $[\omega_0-i\infty,\omega_0+i\infty]$
and along the half-ellipse $C_{\rm L}$, going between $0$ and $\omega_0$ in the lower half-plane.
The advantage of this contour is that one does not need to investigate the pole-structure of the integrand,
which is especially complicated for the diagram $D$.

In order to check the numerical procedure we have performed the calculation
in both Feynman and Coulomb gauges for the photon propagator corresponding
to the interelectronic interaction. The individual terms $A$--$I$
are presented in both gauges for the $g$ factor of $^{208}{\rm Pb}^{79+}$
in Table \ref{tab:contr-g} and for the hyperfine splitting of $^{209}{\rm Bi}^{83+}$
in Table \ref{tab:contr-hfs}. The data for the $g$ factor demonstrate
large cancellation of the individual terms, leading to the loss of 2 digits
in the total value. Due to this fact the uncertainty of our results
for the $g$ factor is significantly larger than the one for the hyperfine splitting.
Moreover, for lower values of $Z$ the convergence of the partial-wave
expansion worsen, and the resulting accuracy becomes unacceptable.
This means that for the $g$~factor in the middle-$Z$ region a special treatment
of the many potential term is required.

%
\section{Two-electron vacuum polarization}
\label{sec:scr-vp}
%

The diagrams, corresponding to the screened vacuum-polarization correction,
are shown in Fig.~\ref{fig:two-el-vp}. Similar to the case of the screened
self-energy we divide the total contribution of these diagrams into the
irreducible and reducible parts. The reducible part should be considered
together with the related contributions from the second and the third
terms in Eq. (\ref{eq:dE-3}).
The irreducible parts of each diagram $A$ -- $F$ are denoted by the same letter,
$\dEScrVPX{A-F}$, while the reducible terms are collected into three groups,
$\dEScrVPX{G,H,I}$.
The total correction due to the screened vacuum-polarization is given by
\be
  \dEScrVP = \sum_{b} \sum_{PQ} (-1)^{P+Q}
    \left( \dEScrVPX{A} + \dEScrVPX{B} + \dEScrVPX{C} + \dEScrVPX{D}
\right.
\nonumber\\
\left.
         + \dEScrVPX{E} + \dEScrVPX{F} + \dEScrVPX{G} + \dEScrVPX{H} + \dEScrVPX{I}
    \right)
\,.
\ee
Here again $P$ and $Q$ are the permutation operators, interchanging $a$ and $b$,
and the summation over $b$ runs over all core electron states.
The notation $\DeltaQbPb\equiv\veps_{Qb}-\veps_{Pb}$ will be used below.

The diagrams $A$, $B$ and $E$ (so-called 'electric-loop' diagrams) involve
the matrix elements of the standard Coulomb-field-induced vacuum-polarization
potential $\UVP$. The charge renormalization makes this potential finite,
and the expressions for the resulting contributions, Uehling and Wichmann-Kroll
potentials, can be found, e.g., in Ref. \cite{mohr:98:prep}.
The contributions of these diagrams are given by
\be
\label{eq:vp-a}
  \dEScrVPX{A} = \dEScrVPX{A1} + \dEScrVPX{A2}
\,,
\ee
\be
\label{eq:vp-a1}
  \dEScrVPX{A1} = 2 \,{\sum_{n_{1,2}}}'
    \matrixel{Pa}{\UVP}{n_1}
    \frac{\matrixel{n_1}{T_0}{n_2}\matrixel{n_2 Pb}{I(\DeltaQbPb)}{Qa Qb}}
         {(\veps_{Pa}-\vepsn{1})(\veps_{Pa}-\vepsn{2})}
\,,
\ee
\be
\label{eq:vp-a2}
  \dEScrVPX{A2} = 2 \,{\sum_{n_{1,2}}}'
    \matrixel{Pa}{\UVP}{n_1}
    \frac{\matrixel{n_1 Pb}{I(\DeltaQbPb)}{n_2 Qb}\matrixel{n_2}{T_0}{Qa}}
         {(\veps_{Pa}-\vepsn{1})(\veps_{Qa}-\vepsn{2})}
\,,
\ee
\be
\label{eq:vp-b}
  \dEScrVPX{B} = 2 \,{\sum_{n_{1,2}}}'
    \frac{\matrixel{Pa}{T_0}{n_1}}{\veps_{Pa}-\vepsn{1}}
    \matrixel{n_2}{\UVP}{n_1}
    \frac{\matrixel{n_2 Pb}{I(\DeltaQbPb)}{Qa Qb}}{\veps_{Pa}-\vepsn{2}}
\,,
\ee
\be
\label{eq:vp-e}
  \dEScrVPX{E} = \dEScrVPX{E1} + \dEScrVPX{E2}
\,,
\ee
\be
\label{eq:vp-e1}
  \dEScrVPX{E1} = 2 \,{\sum_{n_{1,2}}}'
    \matrixel{Pa}{\UVP}{n_1}
    \frac{\matrixel{Pb}{T_0}{n_2}\matrixel{n_1 n_2}{I(\DeltaQbPb)}{Qa Qb}}
         {(\veps_{Pa}-\vepsn{1})(\veps_{Pb}-\vepsn{2})}
\,,
\ee
\be
\label{eq:vp-e2}
  \dEScrVPX{E2} = 2 \,{\sum_{n_{1,2}}}'
    \matrixel{Pa}{\UVP}{n_1}
    \frac{\matrixel{n_1 Pb}{I(\DeltaQbPb)}{Qa n_2}\matrixel{n_2}{T_0}{Qb}}
         {(\veps_{Pa}-\vepsn{1})(\veps_{Qb}-\vepsn{2})}
\,.
\ee
The diagrams of type $C$ (so-called 'magnetic-loop' diagrams) involve
the matrix elements of the magnetic-field-induced vacuum-polarization
potential $\UVPml$.
It also requires the charge renormalization to make it finite.
Our present treatment of this potential is restricted to the Uehling
(free-electron-loop) approximation.
The expression for this potential in the Uehling approximation in case
of the hyperfine interaction can be found in Ref. \cite{artemyev:01:pra}
for the point-dipole nuclear model, and in Ref. \cite{volotka:08:pra}
for the sphere model of the nuclear magnetization distribution.
In case of the $g$ factor this potential in the Uehling approximation is equal to zero.
The contribution of the diagrams of type $C$ is given by
\be
\label{eq:vp-c}
  \dEScrVPX{C} = 2 \,{\sum_{n_{1}}}'
    \matrixel{Pa}{\UVPml}{n_1}
    \frac{\matrixel{n_1 Pb}{I(\DeltaQbPb)}{Qa Qb}}{\veps_{Pa}-\vepsn{1}}
\,.
\ee
The diagram of type $F$ contains the interelectronic-interaction operator
modified by the vacuum polarization $\IVP(\omega)$. Our present treatment
of this term is restricted to the Uehling approximation. The corresponding expression
for $\IVP(\omega)$ in the Feynman gauge can be found, e.g., in Ref. \cite{artemyev:99:pra}.
The contribution of this diagram reads
\be
\label{eq:vp-f}
  \dEScrVPX{F} = 2 \,{\sum_{n_{1}}}'
    \frac{\matrixel{Pa}{T_0}{n_1}}{\veps_{Pa}-\vepsn{1}}
    \matrixel{n_1 Pb}{\IVP(\DeltaQbPb)}{Qa Qb}
\,.
\ee
The contribution of the diagram $D$ is given by the expression
\be
\label{eq:vp-d}
  \dEScrVPX{D} = \matrixel{Pa Pb}{\IVPml(\DeltaQbPb)}{Qa Qb}
\,,
\ee
where the operator $\IVPml$ can be derived within the two-time Green function method.
In the present work we omit this term, assuming its value is relatively small.

The reducible contributions $G$, $H$ and $I$ are given by
\be
\label{eq:vp-g}
  \dEScrVPX{G} = \dEScrVPX{G1} + \dEScrVPX{G2} + \dEScrVPX{G3}
\,,
\ee
\be
\label{eq:vp-g1}
  \dEScrVPX{G1} = - 2 \,{\sum_{n_{1}}}'
    \frac{1}{(\veps_{Pa}-\vepsn{1})^2}
  \Bigl\{
    \matrixel{Pa}{\UVP}{n_1}\matrixel{n_1}{T_0}{Pa}
    \matrixel{Pa Pb}{I(\DeltaQbPb)}{Qa Qb}
\nonumber\\
  + \matrixel{Pa}{\UVP}{Pa}
    \matrixel{Pa}{T_0}{n_1}
    \matrixel{n_1 Pb}{I(\DeltaQbPb)}{Qa Qb}
\nonumber\\
  + \matrixel{Pa}{T_0}{Pa}
    \matrixel{Pa}{\UVP}{n_1}
    \matrixel{n_1 Pb}{I(\DeltaQbPb)}{Qa Qb}
  \Bigr\}
\,,
\ee
\be
\label{eq:vp-g2}
  \dEScrVPX{G2} = 2 \left\{ {\sum_{n_1}}'
    \frac{1}{\veps_{Pa}-\vepsn{1}} \right.
  \Bigl[
    \matrixel{Pa}{\UVP}{n_1}
    \matrixel{n_1}{T_0}{Pa}
    \matrixel{Pa Pb}{I'(\DeltaQbPb)}{Qa Qb}
\nonumber\\
  + \matrixel{Pa}{\UVP}{n_1}
    \matrixel{n_1 Pb}{I'(\DeltaQbPb)}{Qa Qb}
    \Bigl( \matrixel{Qb}{T_0}{Qb} - \matrixel{Pb}{T_0}{Pb} \Bigr)
  \Bigr]
\nonumber\\
  + \matrixel{Pa}{\UVP}{Pa}
    \left[
    {\sum_{n_1}}' \frac{\matrixel{Pa Pb}{I'(\DeltaQbPb)}{n_1 Qb}
                        \matrixel{n_1}{T_0}{Qa}}{\veps_{Qa}-\vepsn{1}}
    \right.
\nonumber\\
  \left.\left.
  + {\sum_{n_1}}' \frac{\matrixel{Pa Pb}{I'(\DeltaQbPb)}{Qa n_1}
                        \matrixel{n_1}{T_0}{Qb}}{\veps_{Qb}-\vepsn{1}}
    \right]
  \right\}
\,,
\ee
\be
\label{eq:vp-g3}
  \dEScrVPX{G3} = \matrixel{Pa}{\UVP}{Pa}
    \matrixel{Pa Pb}{I''(\DeltaQbPb)}{Qa Qb}
    \Bigl( \matrixel{Qb}{T_0}{Qb} - \matrixel{Pb}{T_0}{Pb} \Bigr)
\,,
\ee
\be
\label{eq:vp-h}
  \dEScrVPX{H} = \frac{1}{2}\,\matrixel{Pa Pb}{I'(\DeltaQbPb)}{Qa Qb}
    \Bigl( \matrixel{Qb}{\UVPml}{Qb} - \matrixel{Pb}{\UVPml}{Pb} \Bigr)
\,,
\ee
\be
\label{eq:vp-i}
  \dEScrVPX{I} = \frac{1}{2}\,\matrixel{Pa Pb}{{\IVP}'(\DeltaQbPb)}{Qa Qb}
    \Bigl( \matrixel{Qb}{T_0}{Qb} - \matrixel{Pb}{T_0}{Pb} \Bigr)
\,.
\ee

The numerical calculations of the screened vacuum-polarization corrections
are performed in coordinate space employing the finite basis set
constructed from the DKB-splines \cite{shabaev:04:prl}.
For the electric-loop potential $\UVP$, that enters $\dEScrVPX{A}$, $\dEScrVPX{B}$,
$\dEScrVPX{E}$, and $\dEScrVPX{G}$,
we employ the well-known expression for the Uehling part and the approximate
formulas from Ref. \cite{fainshtein:90:jpb} for the Wichmann-Kroll part.
The magnetic-loop potential $\UVPml$ ($\dEScrVPX{C}$ and $\dEScrVPX{H}$)
is taken in the Uehling approximation only. In case of the $g$ factor this
leads to zero contribution. In case of the hyperfine structure we use
the expression for the extended nucleus from Ref. \cite{volotka:08:pra}.
For the operator $\IVP$ ($\dEScrVPX{F}$ and $\dEScrVPX{I}$) we also
employ the Uehling approximation, the corresponding expressions are taken
from Ref. \cite{artemyev:99:pra}.
The calculational procedure has been checked utilizing the Feynman and Coulomb gauges
for the photon propagator mediating the interelectronic interaction.
The results in both gauges are presented term by term in Tables \ref{tab:contr-g}
and \ref{tab:contr-hfs} for the $g$ factor of $^{208}{\rm Pb}^{79+}$
and for the hyperfine splitting of $^{209}{\rm Bi}^{83+}$, respectively.

%
\section{Results and discussion}
\label{sec:results}
%

\subsection{$\bm{g}$ factor}

The total value of the $g$ factor of a Li-like ion can be written as
\be
\label{eq:gtotal}
  g = \gdirac + \dgint + \dgqed + \dgsqed + \dgnuc
\,.
\ee
Here $\gdirac$ is the Dirac value, $\dgint$ is the interelectronic-interaction correction,
$\dgqed$ is the one-electron QED correction, $\dgsqed$ is the screened QED correction,
and $\dgnuc$ is the contribution of nuclear effects (finite nuclear size, nuclear recoil
and nuclear polarization). To the first order in $\alpha$ and $1/Z$ the screened QED correction
is defined by the set of the two-electron QED diagrams evaluated in the present paper.
The operator $T_0$ in this case reads
\be
  T_0 = \muB \left[\bfr\times\balpha\right]\cdot\bcalH
\,,
\ee
where $\muB = |e|/2$ is the Bohr magneton, $\bcalH$ is the magnetic field
directed along the $z$-axis.
The corresponding contribution to the $g$ factor is given by
\be
\label{eq:dgScrQED}
  \dgsqed = \dgScrSE + \dgScrVP
\,,
\\
  \dgScrSEVP = \dEScrSEVP / ( \muB \calH m_j )
\,,
\ee
where $m_j$ is the $z$-projection of the total angular momentum.
In Table \ref{tab:contr-g} the contributions of the individual terms
to $\dgScrSE$ and $\dgScrVP$ are presented in the Feynman and Coulomb gauges.
The contributions of $\dEScrVPX{C}$ and $\dEScrVPX{H}$ are zero in the Uehling
approximation that we employed for $\UVPml$. The diagram $F$, which is gauge
invariant itself, is calculated only in the Feynman gauge. It can be seen
from the table that the total results in the different gauges are in a fair
agreement with each other.
Additionally, we estimated the Wichmann-Kroll part of the magnetic-loop diagrams
taking the one-electron value from Ref.~\cite{lee:05:pra},
and assuming the same screening ratio as for the electric-loop diagrams.

In Table \ref{tab:total-g} the individual terms and the total values of the $g$ factor
for Li-like lead $^{208}{\rm Pb}^{79+}$ and uranium $^{238}{\rm U}^{89+}$ are presented.
The contribution of the screened self-energy correction $\dgScrSE$ amounts to
$-3.3(2) \times 10^{-6}$ for $^{208}{\rm Pb}^{79+}$
and $-4.9(9) \times 10^{-6}$ for $^{238}{\rm U}^{89+}$.
The estimated uncertainty of the result is rather large due to the cancellation
of the individual terms presented in Table \ref{tab:contr-g}.
The values obtained previously with local screening potentials are
$-3.5(1.2) \times 10^{-6}$ and $-3.1(1.5) \times 10^{-6}$, respectively \cite{glazov:06:pla}.
The contribution of the screened vacuum-polarization $\dgScrVP$ is
$1.53(3) \times 10^{-6}$ for $^{208}{\rm Pb}^{79+}$
and $2.55(5) \times 10^{-6}$ for $^{238}{\rm U}^{89+}$.
The other contributions to the $g$ factor presented in Table \ref{tab:total-g}
were considered in details in our previous studies \cite{glazov:04:pra,glazov:06:pla}.
The accuracy of the $g$ factor value is about $10\%$ better than that
from Ref. \cite{glazov:06:pla} and is almost completely determined
by the higher-orders of the interelectronic-interaction correction.

\subsection{Hyperfine splitting}

The total value of the hyperfine splitting of a Li-like ion can be written as,
\be
  \Delta E^{(a)}_{\rm hfs} = E_F \left[ A(\aZ)(1-\delta)(1-\veps)
    + \frac{1}{Z}B(\aZ) + \frac{1}{Z^2}C(Z,\aZ)
    + \xQED + \xSQED \right]
\,,
\ee
where $E_F$ is the non-relativistic value of the hyperfine splitting (Fermi energy),
$A(\aZ)$ is the one-electron relativistic factor, $\delta$ and $\veps$ are
the corrections for distributions of the charge and magnetic moment over the nucleus,
respectively. The interelectronic-interaction corrections of first and higher orders
in $1/Z$ are represented by the functions $B(\alpha Z)$ and $C(Z,\alpha Z)$, respectively.
The term $\xQED$ corresponds to the one-electron QED corrections.
The details on these contributions can be found in Refs. \cite{shabaev:01:prl,volotka:08:pra}
and references therein.
The operator $T_0$ for the hyperfine splitting is given by
\be
  T_0 = E_F G_a \frac{\left[\bfr\times\balpha\right]_z}{r^3}
\,.
\ee
The factor $G_a$ is defined by the quantum numbers of the valence state,
\be
  G_a = \frac{n^3 (2l+1) j (j+1)}{2 (\aZ)^3 m_j}
\,,
\ee
where $n$ is the principal quantum number, $j$ and $m_j$ are the angular momentum
and its projection, and $l$ defines the parity of the state.
We note that $T_0$ is the effective one-particle operator, which is
employed in calculations of various contributions to the hyperfine splitting.
The full Hamiltonian of the hyperfine interaction is the well-known
Fermi-Breit operator.

To the first order in $\alpha$ and $1/Z$ the screened QED correction $\xSQED$
to the hyperfine splitting is given by
\be
\label{eq:xScrSE}
  \xSQED = \xScrSE + \xScrVP
\,,
\\
  \xScrSEVP = \dEScrSEVP / E_F
\,.
\ee
In Table~\ref{tab:scr-se-hfs} the screened self-energy and vacuum-polarization
corrections to the hyperfine splitting are presented for several values of $Z$
in the range $Z=20$ -- $83$.
The calculations are performed with the Fermi model for the finite nuclear
charge distribution. The finite nuclear magnetization distribution is introduced
via an additional factor $F(r)$ in the operator $T_0$ \cite{volotka:08:pra}.
The individual contributions to $\xScrSE$ and $\xScrVP$ for lithiumlike
bismuth $^{209}{\rm Bi}^{80+}$ are presented
in Table~\ref{tab:contr-hfs} in the Feynman and Coulomb gauges.
Perfect agreement is found between the total results in the different gauges.
This is also true for the other values of $Z$.
We mention, however that the contribution of the screened vacuum-polarization
diagram $F$, which is gauge invariant itself, is calculated in the Feynman gauge only.
For the screened Wichmann-Kroll magnetic-loop part we have employed
the hydrogenic $2s$ value from Ref.~\cite{artemyev:01:pra},
assuming that it enters with the same screening ratio as the Uehling terms.

In Table~\ref{tab:total-hfs} we present the total value of the hyperfine
splitting in bismuth in terms of the specific difference of the
ground state hyperfine splitting in the H-like ion ($1s$) and in the Li-like ion ($2s$):
$\Delta^\prime E = \Delta E^{(2s)} - \xi \Delta E^{(1s)}$.
It was proposed in Ref. \cite{shabaev:01:prl} to consider this difference
in order to overcome the problem of a large uncertainty of the Bohr-Weisskopf (BW) effect,
originating from the nuclear magnetization distribution.
The parameter $\xi$ is chosen to cancel the BW correction,
and the accuracy of the specific difference $\Delta^\prime E$
appears to be much higher than the accuracy of the splittings
$\Delta E^{(1s)}$ and $\Delta E^{(2s)}$ themselves.
The value of $\xi = 0.16886$ has been found for bismuth,
taking into account the BW effect on all of the contributions,
presented in Table~\ref{tab:total-hfs}.
For the related discussion we refer to our Letter \cite{volotka:09:prl}.
We only mention here that the rms radius was taken to be
$\la r^2 \ra^{1/2} = 5.5211$ fm \cite{angeli:04:adndt},
the nuclear spin and parity $I^\pi=9/2-$, and the magnetic moment
$\mu=4.1106(2) \muN$ \cite{stone:05:adndt}.

\subsection{Conclusion}

The rigorous evaluation of the screened QED corrections to the $g$ factor
and to the hyperfine splitting of heavy Li-like ions within \textit{ab initio}
QED approach has been performed. Previously developed procedures
for the evaluation of the one-electron QED corrections in presence
of the external magnetic field and of the two-electron QED corrections
to the energy levels have been associated and generalized.
The complete gauge-invariant set of the two-electron self-energy diagrams
with external magnetic field has been calculated.
The dominant part of the two-electron vacuum-polarization correction
has been calculated as well. The electric-loop diagrams have been evaluated
for both Uehling and Wichmann-Kroll parts.
The magnetic-loop diagrams have been evaluated in the Uehling approximation.
These results improve the accuracy of the theoretical predictions
for the $g$ factor and the hyperfine splitting of heavy ions
where stringent tests of the bound-state QED effects are feasible.

\acknowledgments
Valuable conversations with A.~N.~Artemyev are gratefully acknowledged.
This work was supported by RFBR (Grant No. 07-02-00126-a), GSI, DFG
(Grant No. 436RUS113/950/0-1), and by the Ministry of Education and Science
of Russian Federation (Program for Development of Scientific Potential
of High School, Grant No. 2.1.1/1136; Program ”Scientific and pedagogical
specialists for innovative Russia”) and by the grant of the President
of Russian Federation. D.A.G. acknowledges the support
by the FAIR -- Russia Research Center, and by Saint-Petersburg Government.
V.M.S. acknowledges the support by the Alexander von Humboldt Foundation.
%
%
%

%
%
%
%
%
%
%
\begin{table}
\caption{Contributions of the individual terms to the screened self-energy
and vacuum-polarization corrections to the $g$ factor of Li-like lead $^{208}{\rm Pb}^{79+}$.
The units are $10^{-6}$.
\label{tab:contr-g}}
\vspace{0.5cm}
\begin{tabular}{rr@{}lr@{}lrr@{}lr@{}l}
\hline
\hline
&\multicolumn{4}{c}{Screened SE}
&
&\multicolumn{4}{c}{Screened VP}
\\
& \multicolumn{2}{c}{Feynman}
& \multicolumn{2}{c}{Coulomb}
&
& \multicolumn{2}{c}{Feynman}
& \multicolumn{2}{c}{Coulomb}
\\ \hline
A        &    171.&7 &    173.&6 & A          & $-$34.&27  & $-$34.&65  \\
B        &     18.&1 &     18.&5 & B          &     0.&83  &     0.&82  \\
C        &  $-$30.&5 &  $-$31.&6 & C          &       &--- &       &--- \\
D        &  $-$52.&3 &  $-$52.&0 & D          &       &--- &       &--- \\
E        &   $-$1.&0 &   $-$1.&0 & E          &     0.&16  &     0.&15  \\
F        &      4.&2 &      4.&2 & F          &  $-$0.&01  &       &--- \\
G        & $-$167.&3 & $-$169.&2 & G          &    34.&89  &    35.&29  \\
H        &  $-$41.&3 &  $-$41.&3 & H          &       &--- &       &--- \\
I        &     95.&3 &     95.&5 & I          &     0.&00  &       &--- \\
Total SE &   $-$3.&3 &   $-$3.&3 & Total(A-I) &     1.&60  &     1.&60  \\
         &        &  &        &  & WK-ml      & \multicolumn{4}{c}{$-$0.06(3)}\\
         &        &  &        &  & Total VP   & \multicolumn{4}{c}{\phantom{$-$}1.54(3)}\\
\hline
\hline
\end{tabular}
\end{table}
%
%
\begin{table}
\caption{Contributions of the individual terms to the screened self-energy
and vacuum-polarization corrections to the hyperfine splitting of Li-like bismuth
$^{209}{\rm Bi}^{80+}$ in terms of $\xScrQED$.
\label{tab:contr-hfs}}
\vspace{0.5cm}
\begin{tabular}{rr@{}lr@{}lrr@{}lr@{}l}
\hline
\hline
&\multicolumn{4}{c}{Screened SE}
&
&\multicolumn{4}{c}{Screened VP}
\\
& \multicolumn{2}{c}{Feynman}
& \multicolumn{2}{c}{Coulomb}
&
& \multicolumn{2}{c}{Feynman}
& \multicolumn{2}{c}{Coulomb}
\\ \hline
A        &    0.&001544 &    0.&001555 & A          & $-$0.&0004881 & $-$0.&0004892  \\
B        & $-$0.&000380 & $-$0.&000398 & B          & $-$0.&0002128 & $-$0.&0002103  \\
C        &    0.&001928 &    0.&001952 & C          & $-$0.&0001691 & $-$0.&0001669  \\
D        & $-$0.&000936 & $-$0.&000945 & D          &      &---     &      &---      \\
E        &    0.&000028 &    0.&000028 & E          & $-$0.&0000031 & $-$0.&0000029  \\
F        & $-$0.&000174 & $-$0.&000172 & F          &    0.&0000015 &      &---      \\
G        & $-$0.&001298 & $-$0.&001307 & G          &    0.&0002766 &    0.&0002749  \\
H        &    0.&000331 &    0.&000331 & H          &    0.&0000023 &    0.&0000001  \\
I        &    0.&000066 &    0.&000066 & I          &    0.&0000000 &      &---      \\
Total SE &    0.&001109 &    0.&001109 & Total(A-I) & $-$0.&0005927 & $-$0.&0005927  \\
         &      &       &      &       & WK-ml & \multicolumn{4}{c}{\phantom{$-$}0.00005(2)}\\
         &      &       &      &       & Total VP & \multicolumn{4}{c}{$-$0.00054(2)}\\
\hline
\hline
\end{tabular}
\end{table}
%
%
\begin{table}
\caption{Individual contributions to the ground-state $g$ factor of Li-like
lead $^{208}{\rm Pb}^{79+}$ and uranium $^{238}{\rm U}^{89+}$.
\label{tab:total-g}}
\vspace{0.5cm}
\begin{tabular}{lr@{}lr@{}l}
\hline
\hline
& \multicolumn{2}{c}{$^{208}{\rm Pb}^{79+}$}
& \multicolumn{2}{c}{$^{238}{\rm U}^{89+}$}
\\ \hline
Dirac value (point nucleus)  &    1.&932 002 904      &    1.&910 722 624 (1)  \\
Finite nuclear size          &    0.&000 078 58 (13)  &    0.&000 241 30 (43)  \\
Interelectronic interaction, $\sim 1/Z$
                             &    0.&002 148 29       &    0.&002 509 84       \\
Interelectronic interaction, $\sim 1/Z^2$ and h.o.
                             & $-$0.&000 007 6 (27)   & $-$0.&000 008 5 (38)   \\
QED, $\sim \alpha$           &    0.&002 411 7 (1)    &    0.&002 446 3 (2)    \\
QED, $\sim \alpha^2$         & $-$0.&000 003 6 (5)    & $-$0.&000 003 6 (8)    \\
Screened SE                  & $-$0.&000 003 3 (2)    & $-$0.&000 004 9 (9)    \\
Screened VP                  &    0.&000 001 54 (3)   &    0.&000 002 55 (5)   \\
Nuclear recoil               &    0.&000 000 25 (35)  &    0.&000 000 28 (69)  \\
Nuclear polarization         & $-$0.&000 000 04 (2)   & $-$0.&000 000 27 (14)  \\
Total                        &    1.&936 628 7 (28)   &    1.&915 905 7 (41)   \\
\hline
\hline
\end{tabular}
\end{table}
%
%
\begin{table}
\caption{Screened QED corrections to the hyperfine splitting of lithiumlike ions.
\label{tab:scr-se-hfs}}
\vspace{0.5cm}
\begin{tabular}{lr@{}lr@{}lr@{}l}
\hline
\hline
Z
& \multicolumn{2}{c}{$\xScrSE$}
& \multicolumn{2}{c}{$\xScrVP$}
& \multicolumn{2}{c}{$\xScrQED$}
\\ \hline
20 &  0.&000256  & $-$0.&000116    & 0.&000140(1) \\
30 &  0.&000330  & $-$0.&000131    & 0.&000199(1) \\
40 &  0.&000394  & $-$0.&000155    & 0.&000238(1) \\
50 &  0.&000473  & $-$0.&000186(3) & 0.&000287(3) \\
60 &  0.&000582  & $-$0.&000241(4) & 0.&000340(4) \\
70 &  0.&00075   & $-$0.&00033(1)  & 0.&00042(1)  \\
83 &  0.&00111   & $-$0.&00054(2)  & 0.&00057(2)  \\
\hline
\hline
\end{tabular}
\end{table}
%
%
\begin{table}
\caption{Individual contributions to the specific difference $\Delta^\prime E$
of the hyperfine splittings for bismuth $^{209}$Bi. The units are meV.}
\label{tab:total-hfs}
\begin{tabular}{lr@{}lr@{}lr@{}l} \hline\hline
  & \multicolumn{2}{c}{$\Delta E^{(2s)}$} & \multicolumn{2}{c}{$\xi \Delta E^{(1s)}$}
           & \multicolumn{2}{c}{$\Delta^\prime E$}                \\ \hline
Dirac value           &   844.&829    &  876.&638 & $-$31.&809    \\
Interelectronic interaction, $\sim 1/Z$
                      & $-$29.&995    &      &    & $-$29.&995    \\
Interelectronic interaction, $\sim 1/Z^2$ and h.o.
                      &     0.&25(4)  &      &    &     0.&25(4)  \\
QED                   & $-$ 5.&052    & $-$5.&088 &     0.&036    \\
Screened SE           &     0.&381    &      &    &     0.&381    \\
Screened VP           & $-$ 0.&187(6) &      &    & $-$ 0.&187(6) \\
Total                 &       &       &      &    & $-$61.&32(4)  \\
\hline\hline
\end{tabular}
\end{table}
\end{document}